\documentclass[doublecol]{epl2} 
\usepackage{amsmath}
\usepackage{times}
\usepackage{amssymb}
\usepackage{graphicx}
\usepackage{float}
\usepackage{color}

\title{
Universal current fluctuations in    the symmetric exclusion process and other  diffusive systems
}
\shorttitle{Universal current fluctuations in exclusion processes
} 
 \author{Eric Akkermans \inst{1} \and  Thierry Bodineau  \inst{2} \and Bernard Derrida \inst{3}
 \and Ohad Shpielberg   \inst{1} }
\institute{ 
\inst{1} 
Department of Physics, Technion, Israel Institute of Technology, 32000  Haifa, Israel
\\ 
\inst{2}
 D\'epartement de Math\'ematiques et Applications, 
 Ecole Normale Sup\'erieure,
45 rue d’Ulm, 
75230 Paris cedex 05, France
\\ 
\inst{3}  Laboratoire de Physique Statistique, Ecole Normale
Sup\'erieure, UPMC Paris 6, Universit\'e Paris Diderot, CNRS,
24 rue Lhomond, 75231 Paris cedex 05, France}

  {Nonequilibrium and irreversible thermodynamics}
\abstract{
We  show, using the macroscopic fluctuation theory 
of Bertini, De~Sole, Gabrielli, Jona-Lasinio, and Landim, 
  that the statistics of the current  of the symmetric simple exclusion  process (SSEP) connected to two reservoirs  are the same on an arbitrary  large finite domain  in dimension $d$ 
as in the one dimensional case. Numerical results on squares support  this claim while results on cubes exhibit some  discrepancy.
We argue that  the results of the macroscopic fluctuation theory should be recovered by increasing the size of the contacts.
The generalization to other diffusive systems is straightforward. 
}

\begin{document}

\maketitle


\section{Introduction}
When a system is  connected  for a long time to two heat baths at unequal temperatures or to two reservoirs of particles at different densities, it reaches a non-equilibrium steady state, with a non vanishing current of heat or of particles. This current fluctuates with time and the study of its fluctuations and of its large deviations  has become a central aspect in the theory of non equilibrium systems \cite{PS2,BDGJL5,BDGJL6,derrida2007,HRS1,Kundu,LM,LT,GKP,GLMV,HG,HG1,PM2}. 
A quantity which characterizes these fluctuations is the probability $P(Q_t)$  of observing an energy  or a number of particles $Q_t$
flowing through
the system during a time window $t$. 
A notorious property of   these fluctuations is  known as the fluctuation theorem \cite{ECM,GC,LS,Harris-S} which establishes a general  relation between $P(Q_t)$ and $P(-Q_t)$, starting from   some time reversal symmetry of the microscopic dynamics.
Apart from simple models, however, it  is usually difficult to predict the whole shape of the distribution $P(Q_t)$.

The one-dimensional SSEP (symmetric simple exclusion process)  describes a 
 chain
of $L$ sites on which particles diffuse with hard core repulsion on the same site, connected  at its two ends to two reservoirs at densities $\rho_a$ and $\rho_b$.  It is one of the few examples for which it has been possible to determine the  distribution $P(Q_t)$ both by microscopic and macroscopic approaches \cite{DDR,BD2004,BDGJL5,BDGJL6}.   This distribution reveals that, for a long chain  of length $L$ connected at its two ends to two reservoirs at densities $\rho_a=1$ and  $\rho_b=0$, the Fano factor (the ratio of the first two cumulants) is given,
in the long time limit, by 
\begin{equation}
\lim_{L \to \infty} \ \lim_{t \to \infty} {\langle  Q_t^2  \rangle - \langle  Q_t  \rangle^2 \over \langle  Q_t  \rangle} = {1 \over 3}  \ . 
\label{un-tiers}
\end{equation}
In fact for the SSEP in one dimension, all the cumulants of the current are known \cite{DDR,BD2004} for arbitrary densities $\rho_a$ and $\rho_b$ of the left and of the right reservoirs,  with their generating function given by
\begin{equation}
 \lim_{L \to \infty} \ \lim_{t \to \infty} { L \over t}   { \log \left\langle  \exp[\lambda Q_t]  \right\rangle } = 
\Big( {\rm Arcsinh} (\sqrt{\omega}) \Big)^2
\label{generating-function}
\end{equation}
where $\omega$ is given by
\begin{equation}
\omega= \rho_a (e^\lambda -1) + \rho_b (e^{-\lambda} -1) -  \rho_a (e^\lambda -1)  \rho_b (e^{-\lambda} -1) \ . 
\label{omega-def}
\end{equation}
Note as a striking fact  \cite{DDR,BD2004} that the generating function  which is in principle a function of the three variables $\rho_a, \rho_b$ and $\lambda$ turns out to be a function of the single variable $\omega$.

The cumulants which  can be determined  from (\ref{generating-function},\ref{omega-def}) are the same as those which have been computed for free fermions transmitted through multichannel disordered conductors \cite{BB1,BS,LLY}.  
Although most of the theoretical approaches in the classical (SSEP) case and in the quantum case are  different, the description of  the quantum problem based on  the Boltzmann Langevin    approach has  similarities \cite{Nagaev,SL,PJSB,KvO}
with the macroscopic fluctuation  theory  \cite{BDGJL3,BDGJL5,BDGJL6,BD2005} which is the appropriate theory  to describe diffusive systems such as the SSEP on large scales. 
These similarities rely on the fact  that  the hard core repulsion of  the   exclusion  process  mimics { properly  the Pauli principle for fermions in the quantum problem \cite{BB1}}. In the quantum case, it has been shown  \cite{SL} that the Fano factor takes the value $1/3$ for arbitrary geometries. 

One can  wonder how the cumulants of the integrated current are modified in higher dimension ($d > 1$) and for 
more complicated graphs than linear chains.
A   numerical study   of  the Sierpinski gasket  \cite{GTB},
 with  two corners  connected to reservoirs at densities $\rho_a=1$ and $\rho_b=0$,
 has shown that in this case too, the Fano factor  is still  equal to $1/3$ as in one dimension (\ref{un-tiers}).
The question addressed in this letter is the degree of generality of (\ref{un-tiers}) or (\ref{generating-function},\ref{omega-def}).

It is known  \cite{DG}
that  for an arbitrary finite graph connected to two reservoirs at densities $\rho_a$ and $\rho_b$, all the dependence of the generating function  of the cumulants ${ \log \left\langle  \exp[\lambda Q_t]  \right\rangle }$  on $\lambda, \rho_a$ and $\rho_b$ is    always  through  the single parameter $\omega$  defined in (\ref{omega-def}). This is  proven \cite{DG}
  at an arbitrary time  $t$,  by choosing the  configuration  at $t=0$  of the SSEP with  the right measure. In the long time limit, this   becomes  true for an arbitrary choice of initial conditions.
Therefore for an arbitrary graph one knows in advance that
\begin{equation}
 \lim_{t \to \infty} {1 \over t}   { \log \left\langle  \exp[\lambda Q_t]  \right\rangle } = 
G(\omega)
\label{generating-function-general}
\end{equation}
where $G$ is a function  which depends  on the characteristics of  the considered graph, on  the way it is connected to the reservoirs and on the parameter $\omega$ defined in  (\ref{omega-def}).

\begin{figure}[t]
  \ \  \includegraphics[width=7.5cm]{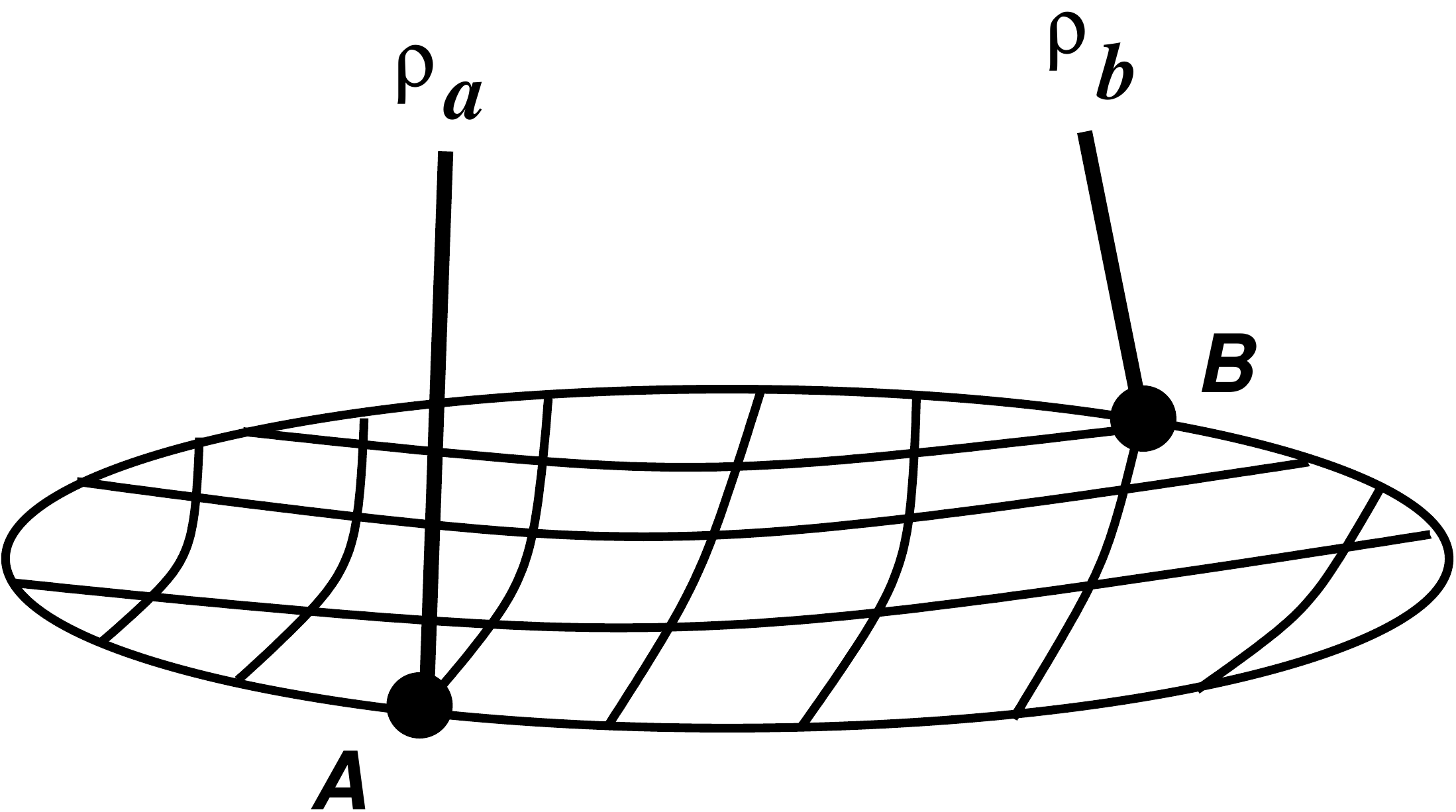}
  \caption{Finite graph with sites $A$ and $B$ maintained at densities $\rho_a$ and $\rho_b$ by reservoirs. }
\label{fig1}
\end{figure}

One purpose of this letter is to show that  for  a large class of graphs and geometries such as represented in figure  \ref{fig1}, when the distance $L$ between the contacts is large, the function $G(\omega)$  is the same as  for a linear chain, up to a multiplicative factor  $\kappa(L)$, namely,
\begin{equation}
G(\omega)   \simeq \kappa(L)
\Big( {\rm Arcsinh} (\sqrt{\omega}) \Big)^2 \  .
\label{generating-function-general-bis}
\end{equation}
In (\ref{generating-function-general-bis}), all the dependence  on the  shape of the graph, on the location of the points $A$ and $B$ connected to the reservoirs, on the nature of the connections,  and on the system size are encoded in the factor   $\kappa(L)$. As a consequence,  for all $\rho_a$ and $\rho_b$, the ratio bewteen any pair of cumulants of $Q_t$ is exactly the same as for the linear chain.

The term  linear in $\lambda$ in (\ref{omega-def},\ref{generating-function-general},\ref{generating-function-general-bis}) gives the  relation between the average current and the factor $\kappa(L)$ in (\ref{generating-function-general-bis})
$$ \lim_{t \to \infty} {\langle Q_t \rangle  \over t}  = \kappa(L) (\rho_a- \rho_b)  \  .  $$

Our approach is based on the macroscopic fluctuation theory \cite{BDGJL3,BDGJL5,BDGJL6,BDGJL7,BD2005}  which requires solving non-linear  differential equations for the current and density profiles which produce  a certain $Q_t$. As the macroscopic fluctuation theory is formulated in the continuum, one needs the density and current profiles to vary slowly on the graph. Therefore our main assumption is that the graph is large, that  one can define derivatives with respect to space on this graph, and that  the contacts  $A$ and $B$  with the reservoirs are large enough to avoid singular gradients of density or current profiles near these contacts.

\section{Computation on finite graphs}
\begin{figure}[t]
  \ \  \includegraphics[width=7.5cm]{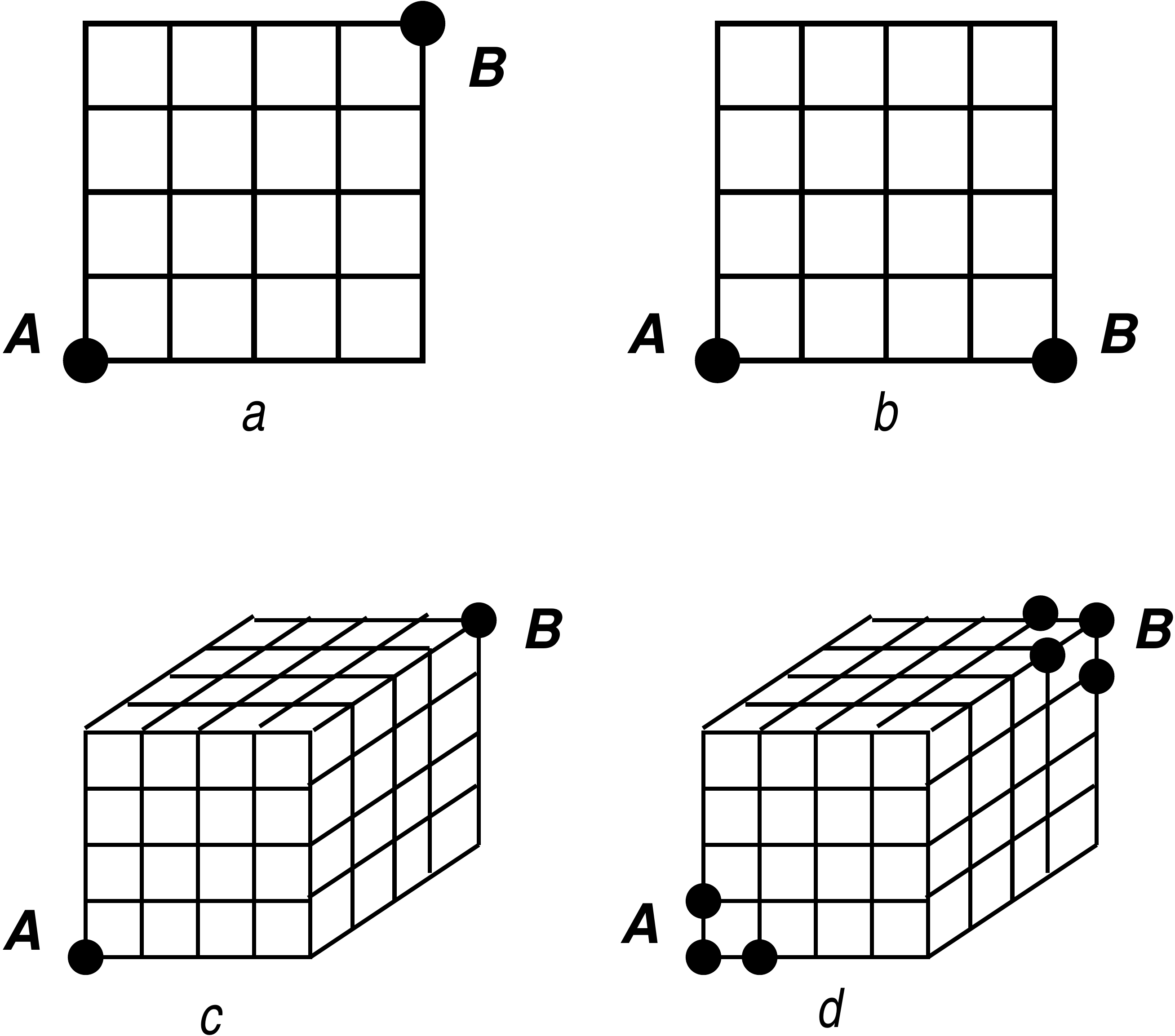}
  \caption{The Fano factors have been determined numerically for squares and cubes of linear size $L$.
In each case the black dots indicate  the points of the lattice which are maintained at density $1$ (sites $A$) and $0$ (sites $B$) by the reservoirs.  A single site is connected to each reservoir in cases {\it a,b,c} while in case {\it d}, there are 4 sites connected to each reservoir.
 }
\label{squa-cub}
\end{figure}

Let us consider the SSEP on  a {\it finite} graph. Each site $i$ on this graph is connected to its neighboring sites  on the graph  (this number  of neighbors  may depend on $i$).
In addition,  two particular sites (or two particular sets of points as in  figure \ref{squa-cub}) are connected to reservoirs as in figure \ref{fig1}:  site $i=A$  is connected  to a reservoir at density $\rho_a$ and site $B$ is connected to a reservoir at density $\rho_b$. At a given time $t$ each site $i$  is either occupied ($\tau_i(t)=1$) or empty ($\tau_i(t)=0$). By definition of the dynamics, the occupation numbers of each pair of connected sites on the lattice are exchanged at rate $1$.
Moreover,  the   effect of reservoirs can be  represented by the rates at which particles are injected on sites $A$ and $B$ when these sites are empty and the rates at which particles are removed from these two sites, when they        are occupied. 
One simple way of updating   sites $A$ and $B$ is  for example  to   keep at any time $t$, $\tau_A(t)= 1$ with probability $\rho_a$ and $\tau_A(t)=0$ with probability $1- \rho_a$, (and  similar probabilities for site $B$) irrespective of the past history of the lattice. With this prescription, the occupation numbers of sites $A$ and $B$ are not correlated to the other sites  of the lattice, they have no correlation at unequal times and  on average one has
$\langle \tau_A(t) \rangle =\rho_a $ and $ \langle \tau_B (t) \rangle =\rho_b $.

Given some random initial condition, the evolution
 of the average density $\rho_i = \langle \tau_i(t) \rangle  $  on site $i$ is given
 for all $i \neq A,B$  by 
\begin{equation}
{d \rho_i \over dt} = \sum_{j \sim  i} \rho_j - \rho_i
\label{heat1}
\end{equation}
where the sum is over the neighbors $j$  of site $i$ on the graph. 
At sites $A$ and $B$, one has $\rho_A= \rho_a$ and $\rho_B= \rho_b$.
 The  average profile $\bar{\rho_i}$, in the non equilibrium steady state,
 satisfies everywhere the  lattice version of the Laplace equation
\begin{equation}
 \Delta \bar{\rho_i}  \ \equiv  \  \sum_{j \sim  i} \bar\rho_j - \bar\rho_i  \  = \ 0 \
\label{Laplacian}
\end{equation}
except in $i=A$ and $i=B$. Denoting 
$Q_t$ the number of particles  flowing through the system (see just below for a precise definition)  during time $t$, our goal is to determine $\mu(\lambda)$
defined by
\begin{equation}
\left\langle e^{\lambda  \, Q_t } \right\rangle \sim e^{\, \mu(\lambda) \, t} \ \ \ \ {\rm for \ large \ } t  \ . 
\label{mu-def}
\end{equation}
 The knowledge of $\mu(\lambda)$ allows to determine  all the cumulants of $Q_t$
$$ \lim_ {t \to \infty}{ \langle Q_t ^n \rangle \over t} = \left.  {d^n \mu(\lambda )\over d \lambda^n } \right|_{ \lambda=0} \ .
$$

We now need to define precisely the number $Q_t$ of particles flowing through the system.
One could count the number $Q_t^{A}$ of particles    flowing  between the reservoir at density $\rho_a$ and  site $A$ ({\it i.e.} the total number of particles injected in site $A$). 
One could alternatively count the number $Q_t^{B}$  of particles flowing  from site $B$ to the reservoir at density $\rho_b$.
Here, based on \cite{BDL2008}, we choose the following  definition of  $Q_t$
\begin{equation}
Q_t = {\frac{1}{2}} \sum_{i,j} (   V_i-    V_j)\  q_{i,j}(t) 
\label{counting}
\end{equation}
where $q_{i,j}(t)= - q_{j,i}(t)$ is the number of particles transferred from site $i$ to site $j$ during time $t$
and  $V_i$ is an arbitrary function defined on each site $i$ 
except for $V_A$ and $V_B$ which are given by 
\begin{equation}
 V_A = 1 \ \ \ 
\ ; \ \ \ \  V_B=0
\label{VAVB}
\end{equation}
 and the factor $1/2$ results from double counting of each bond.

Using that $ \tau_i(0)- \tau_i(t)=\sum_{j \sim i} q_{i,j}(t) $
(where $\tau_i(t)$ is the number of particles at site $i$ and time $t$),
one can rewrite 
(\ref{counting}) as
\begin{eqnarray*}
Q_t =    \sum_i V_i  \sum_{j \sim i} q_{i,j}(t) 
=        V_A \sum_{j \sim A} q_{A,j}(t) +  \ \ \ \ \  \ \ \ \ \ \
 \\  \nonumber 
 \ \ \ \ \  \ \ \ \ \       V_B \sum_{j \sim B} q_{B,j}(t) 
   +   \sum_{i \not = A,B}    V_i \big( \tau_i (0) -\tau_i(t) \big)    
\end{eqnarray*}
 which, using  (\ref{VAVB}), becomes
\begin{equation}
Q_t =    
 \sum_{j \sim A} q_{A,j}(t) + 
      \sum_{i \not = A,B}    V_i \big( \tau_i (0) -\tau_i(t) \big)    
\label{counting-bis}
\end{equation}
Since the graph is  finite,  particles cannot accumulate on the graph (because the number of particles cannot exceed the number $N$ of lattice sites).  
Clearly the first  term in the {\it r.h.s.} of (\ref{counting-bis}) can become arbitrary large with time while the second term remains bounded ($< \sum_i |V_i|$). Therefore
changing the   $V_i$'s changes $Q_t$ by an  amount which cannnot grow with time, so that the value of $\mu(\lambda)$ in (\ref{mu-def})  is the same for any choice of the $V_i$'s. Similarly one can show that the differences $Q_t - Q_t^A$ or $Q_t-Q_t^B$   remain bounded in time
(for example by choosing all $V_i=0$ for $i \neq A$, one has
$Q_t-Q_t^A= \tau_A(0) - \tau_A(t)$ which obviously remains bounded).
Therefore all the definitions of $Q_t$ one can think of lead 
in the long time limit to the same $\mu(\lambda)$.

Here we find convenient to take for $V_i$ a solution of the Laplace equation
\begin{equation}
 \Delta         V_i  \ \equiv  \  \sum_{j \sim  i}        V_j -        V_i  \  = \ 0 \ 
\label{Laplacian-V}
\end{equation}
with boundary conditions \eqref{VAVB}.
(Thus  when $\rho_a \neq \rho_b$, one has simply $V_i=(\rho_i - \rho_b)/(\rho_a-\rho_b)$).

Using a method similar to the one used in  \cite{GTB} (the method consists  essentially in calculating the steady state densities and  two point correlations
as in \cite{DDR}) we have determined numerically the  Fano factor 
$$F= \lim_{t \to \infty}  {\langle  Q_t^2 \rangle - \langle Q_t \rangle^2 \over \langle Q_t \rangle}$$
for  finite squares  of size $L \times L$  and  cubes of size $L \times L \times L$ with open boundary conditions.
The systems are connected to the reservoirs  at densities $\rho_a=1$ and $\rho_b=0$ as in figure  \ref{squa-cub}.

\begin{figure}
  \ \  \includegraphics[width=8.5cm]{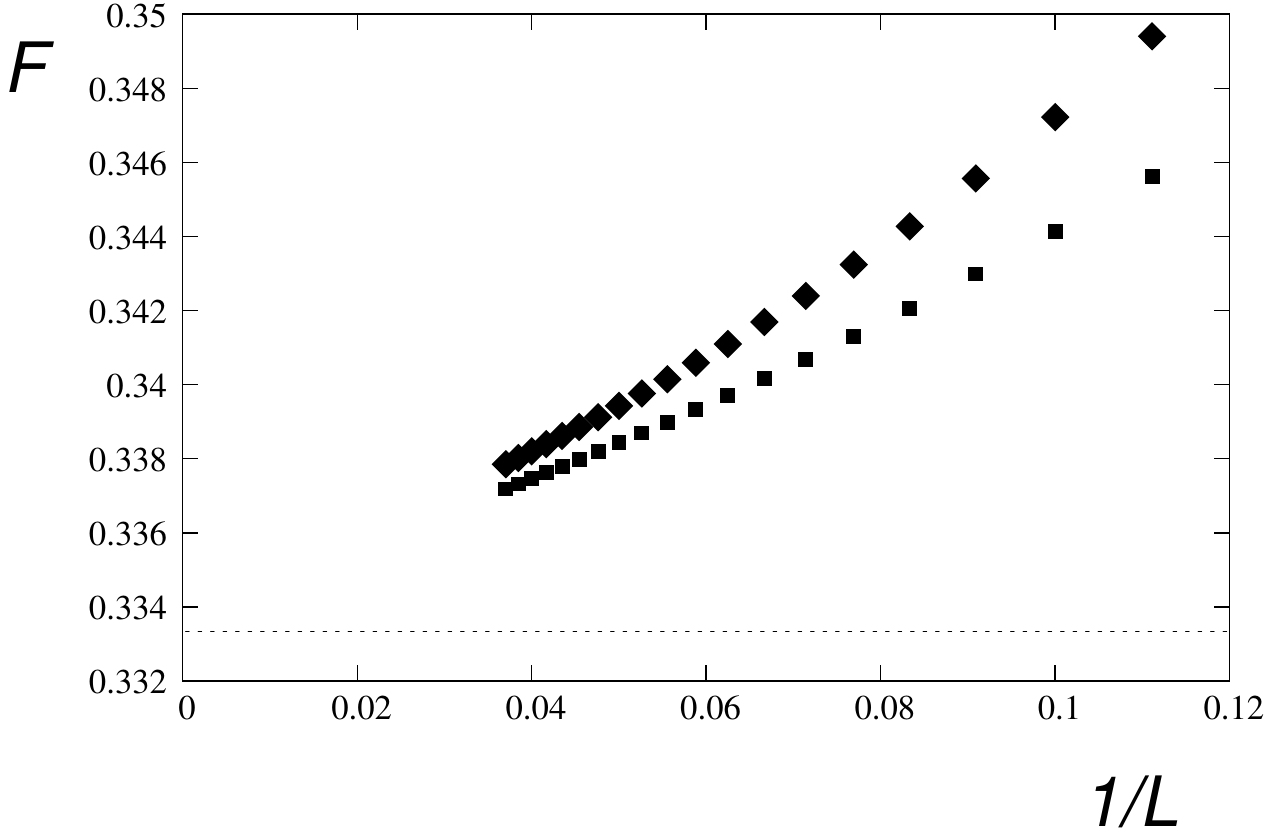}
  \caption{The Fano factor as a function of  $1/L$ is displayed for the two geometries (see figure \ref{squa-cub}) {\it a} (diamonds) and {\it b} (squares) of squares of size up to $27 \times 27$. The results indicate convergence to $1/3$
 for both geometries. }
\label{carre}
\end{figure}

 The results   are shown in figure \ref{carre}  for the {square}.
For both geometries  {\it a} and {\it b} of figure \ref{squa-cub}, the results seem to  converge to $F=1/3$ as $L \to \infty$. 
In figure \ref{carre}   the results are plotted versus $1/L$. We have no particular  reason  
to think that this is the right convergence law. Trying other power law extrapolations show equally well the convergence to $1/3$.

\begin{figure}
  \ \  \includegraphics[width=8.5cm]{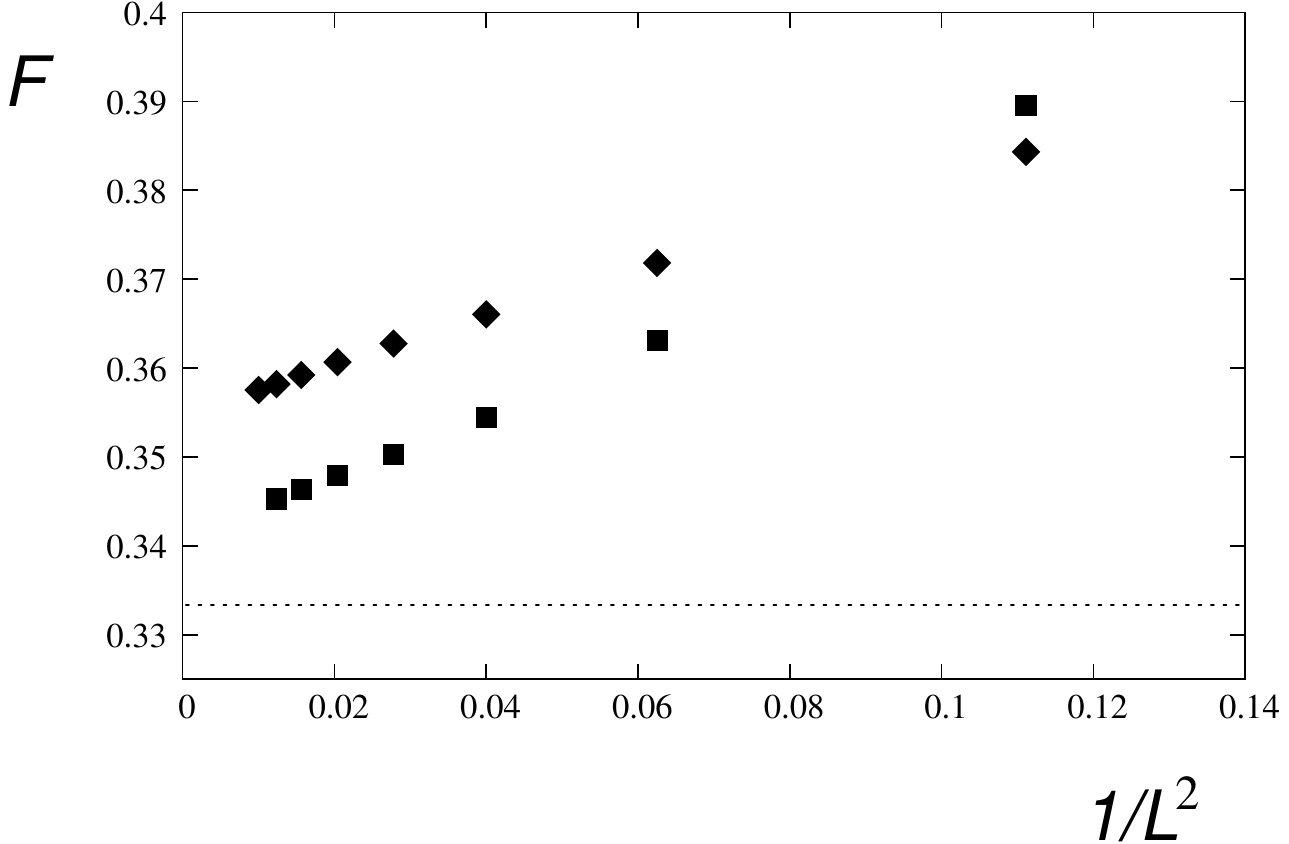}
  \caption{The Fano factor as a function of $1/L^2$ is displayed for the two geometries (see figure \ref{squa-cub})  {\it c} (diamonds) and {\it d} (squares) of cubes  of size up to $10 \times 10 \times  10$. The results  do not seem  to converge to $1/3$, but for the geometry {\it d} the extrapolated value is closer to 1/3. As the size of the contacts increases, keeping it small compared to the system size, the Fano factor  becomes  closer to $1/3$.
 }
\label{cube}
\end{figure}

In the case of the cube,  our data in figure  \ref{cube} indicate a large $L$ limit distinct from $1/3$ for both geometries {\it c} and {\it d} of figure  \ref{squa-cub}.
For geometry {\it d}, however, where the contacts  with the reservoirs are larger, the limiting value is closer to $1/3$.
We will argue below, that if the size of the contacts is macroscopic (for example if the region in contact with each reservoir  contains $(\epsilon L)^d$ sites, no matter how small $\epsilon$  is provided that it  remains finite as $L \to \infty$), the Fano factor should be $1/3$ in any dimension and for all geometries, in agreement with the trend we see going from geometry {\it c} to geometry {\it d} in figure \ref{cube}.

\section{The macroscopic fluctuation theory}
 We consider now a continuous theory, on a $d$-dimensional domain of arbitrary  shape.
The typical size of the domain is $L$ (with $L$ large compared to the  distance between any pair of  connected sites on the lattice). Each reservoir maintains at a fixed  density 
( respectively $ \rho_a$ and $\rho_b$  for reservoirs $A$ and $B$) all the sites at a distance less than $L \epsilon$  from the center of the connection. Here $\epsilon$  is small but remains fixed as $L$ increases, so  the number of sites  in contact with the reservoirs increases with  $L$  but  remains a small fraction of the total number of sites in the graph.

If we introduce a macroscopic coordinate $\vec{r}$ on the domain, such that $|\vec{r_A} - \vec{r_B} | = O(1)$ (so that there is a rescaling by a factor $L$ between the  microscopic cooordinates  on the graph and the  macroscopic 
coordinate $\vec{r}$),
 the densities $\rho_i$ become a scalar  field  $\rho(\vec{r})$ whose time evolution is  the continuous version of (\ref{heat1})
$${d \rho \over dt} = {1 \over L^2} \Delta \rho$$
showing that there is  a  rescaling factor  $L^2 $ between the microscopic time $t$ and the macroscopic time  $\tau = t/L^2$. 
Therefore on a macroscopic  scale the density $\rho(\vec{r},\tau)$ is a function of the macroscopic variables $\vec{r}$ and $\tau$ with the boundary conditions
\begin{equation}
\rho( \partial A,\tau)=\rho_a  \ \ \ {\rm and } \ \ \ \ \rho( \partial B,\tau )= \rho_b 
\label{boundary}
\end{equation}
at the contacts $\partial A$ and $\partial B$ with the reservoirs.

Associated with  the density $\rho(\vec{r},\tau)$ the current   $\vec{j}(\vec{r},\tau)$ is a vector field which satisfies the conservation law
\begin{equation}
{d \rho(\vec{r},\tau) \over d \tau }+ \vec{\nabla} \, .\,  \vec{j}(\vec{r},\tau) =0 \ .
 \label{conservation-1}
\end{equation}

The  potential $v(\vec{r})$, which   is the continuous version of the potential $V_i$  introduced in (\ref{counting}) to measure the current,
  satisfies (see (\ref{Laplacian-V})) 
\begin{equation}
\Delta v(\vec{r}) =0
\ \ \ 
\label{Laplacian-1}
\end{equation}
 with
\begin{equation}
  v( \partial A)=1 \  \ \  {\rm and} \ \  \  v(\partial B )=0
\label{bound-cond}
\end{equation}
 and Neumann boundary conditions otherwise. 

Then  (\ref{counting}) becomes
\begin{equation}
Q_t  = -  L^d \int_0^{t \over L^2} d\tau \ \int d \vec{r} \  \   \vec{\nabla} v(\vec{r}) \,  . \, \vec{j}(\vec{r},\tau)
\label{counting-2}
\end{equation}
where the factor $L^d$  comes from the  change of scale.

According to the macroscopic fluctuation theory \cite{BDGJL3,BDGJL5,BDGJL6,BDGJL7}, the probability of obtaining  time dependent profiles for both density and current over a long time interval $t$, is 
$${\rm Pro} (\{\rho(\vec{r},\tau),\vec{j}(\vec{r},\tau) \}) \sim \exp\left[- L^d I_t \right]$$
with the action  $I_t$ given by
\begin{equation}
\label{action}
I_t = 
 \int_0^{t / L^2} d\tau     { [\,\vec{j}(\vec{r},\tau) +  D( \rho(\vec{r},\tau)) \ \vec{\nabla}\rho(\vec{r},\tau)\, ]^2 \over 2 \sigma(\rho(\vec{r},\tau) )}   \ .
\end{equation}
 This action simply  means  that  the current is  the sum of a Fick's law contribution 
$-D(\rho) \nabla \rho$   and   of a Gaussian noise
$\vec{\eta}$ of variance $\sigma$,     $\delta$-correlated both in time and space 
$$\vec{j}(\vec{r},\tau)= -   D( \rho(\vec{r},\tau))  \ \vec{\nabla}\rho(\vec{r},\tau) + \vec{\eta}(\vec{r},\tau)  \ . $$ 
 A diffusive system is characterized by  the two functions $D(\rho)$ and $\sigma(\rho)$ which show up in (\ref{action}). For the SSEP, one has  \cite{BD2004}
\begin{equation}
D(\rho)=1 \ \ \ ; \ \ \ \sigma(\rho)= 2 \rho(1-\rho) \ . 
\label{D-sigma-SSEP}
\end{equation}
Here we keep $D$ and $\sigma$ arbitrary as our calculation  below is  valid for more  general diffusive systems.
Then, the generating function $\mu(\lambda)$ defined in (\ref{mu-def}) is 
\begin{equation}
\label{mu-1}
\mu(\lambda)= \lim_{t \to \infty} \ 
 \max_{\{\vec{j},\rho\} } 
\left[  {\lambda Q_t - L^d I_t \over t } \right]
\end{equation}
 with $Q_t$ and $I_t$ given by (\ref{counting-2},\ref{action}) and 
where we have  to optimize over all  the time dependent density and current profiles $\{\vec{j}(\vec{r},\tau),\rho(\vec{r},\tau)\}$  which satisfy the conservation law (\ref{conservation-1}).

In what follows, we will assume,  as with the additivity principle \cite{BD2004,BD2005,HG1}, that the optimum in (\ref{mu-1}) is achieved by {\it time independent density and current profiles} 
 $\{\vec{j}(\vec{r}),\rho(\vec{r})\}$. In \cite{BDGJL5,BDGJL6}  it was proven that this assumption is valid for the SSEP
as well as for  a large class of  other diffusive systems (see  \cite{BDGJL5,BDGJL6} for precise conditions on $D$ and $\sigma$).
Then (\ref{mu-1}) can be rewritten
\begin{eqnarray}
\label{mu-2} 
\mu(\lambda)
=  - L^{d-2}
 \min_{\{\vec{j}(\vec{r}),\rho(\vec{r}) \} } 
 \int d \vec{r} \  \  \Big(  \lambda \vec{\nabla} v(\vec{r}) \,  . \, \vec{j}(\vec{r}) \ 
 \\ 
   \ \ \ \ \ + \  { [\, \vec{j}(\vec{r}) +  D(\rho(\vec{r})) \ \vec{\nabla}\rho(\vec{r}) \, ]^2 \over 2 \sigma(\rho(\vec{r})) } \Big)
\nonumber 
\end{eqnarray}
and the conservation law (\ref{conservation-1}) becomes
\begin{equation}
 \vec{\nabla} \, .\,  \vec{j}(\vec{r}) =0 \  . 
 \label{conservation-2}
\end{equation}

By optimizing (\ref{mu-2}) over $\rho(\vec{r})$  
and  $ \vec{j}(\vec{r})$, given the conservation law (\ref{conservation-2}), we deduce,
   \begin{eqnarray}
\nonumber
   D'(\rho) \  { (\vec{j} + D(\rho) \  \vec{\nabla}\rho). \vec{\nabla} \rho  \over \sigma } -
   \sigma'(\rho )  \ 
  { [\vec{j} + D(\rho) \  \vec{\nabla}\rho ]^2 
  \over 2 \sigma(\rho )^2 }
   \\ -  \vec{\nabla} \, . \, \left( D(\rho) \     {\vec{j} +  D(\rho) \ \vec{\nabla}\rho  \over  \sigma(\rho )} \right)= 0
   \label{eq-2}
  \end{eqnarray}
  \begin{equation}
  \label{courant}
  \vec{j}=
  -  D(\rho) \vec{\nabla}\rho  + \sigma(\rho ) \left( \vec{\nabla} h - \lambda   \vec{\nabla}  v \right) 
  \end{equation}
where $\rho,v,\vec{j} $ and $h$ are functions of $\vec{r}$.
 (Here $ h(\vec{r})$ is a Lagrange multiplier associated to the constraint  (\ref{conservation-2}).

Since the conservation law  (\ref{conservation-2}) is not satisfied at the contacts  $A$ and $B$,  there is no
Lagrange parameters at the boundaries and 
\begin{equation}
h( \partial A)=h( \partial B)=0 \ .
\label{hahb}
\end{equation}

Upon defining a function $H(\vec{r})$ by 
$$H(\vec{r})= h(\vec{r}) -  \lambda \,  v(\vec{r}) $$
instead of $h(\vec{r})$, we see that (\ref{courant}) and (\ref{hahb}) become
\begin{equation}
\label{courant-2}
\vec{j}(\vec{r})=
-  D(\rho(\vec{r})) \, \vec{\nabla}\rho(\vec{r})  + \sigma(\rho(\vec{r}) )  \vec{\nabla} H(\vec{r})  
\end{equation}
 and
\begin{equation}
H( \partial A)= - \lambda \ \ \ \ \ \ ; \ \ \ \ \ H( \partial B)=0 \, \, .
\label{HaHb}
\end{equation}
Inserting the expression (\ref{courant-2}) of $\vec{j}(\vec{r})$ into (\ref{conservation-2}) and (\ref{eq-2}), we deduce that $H$ and $\rho$  should satisfy
\begin{equation}
\label{eq-1}
\vec{\nabla} . \Big( D(\rho(\vec{r}))   \,  \vec{\nabla}  \rho(\vec{r}) \Big)=  \vec{\nabla}   .  \left(    \sigma(\rho(\vec{r}) ) \, \vec{\nabla} H(\vec{r})\right)
\end{equation}
and 
\begin{equation}
\label{eq-3}
D(\rho (\vec{r})) \ \Delta H(\vec{r})=  -  {\sigma'(\rho(\vec{r}) )  \over 2}  \left(\vec{\nabla} H(\vec{r})\right)^2 \, \, .
\end{equation}
So the problem of computing $\mu(\lambda)$ in (\ref{mu-2}) is reduced to finding  $H$ and $\rho$ solutions of (\ref{eq-1},\ref{eq-3}) with the boundary conditions (\ref{HaHb})
\begin{equation}
\label{eq: bords rho}
\rho( \partial A) = \rho_a \ \ \ \  , \ \ \ \ 
\rho(\partial B) =  \rho_b 
\end{equation}
and $\vec{j}(\vec{r})$ given by (\ref{courant-2}).

\section{  The link between the one dimensional case and higher dimensions}
We wish now to argue that if one knows the solutions of (\ref{eq-1},\ref{eq-3}) in one dimension, then one can obtain  the solutions for an  arbitrary  domain in any dimension.
 Consider the case of a one dimensional chain of length $L$ in contact with reservoirs at its boundary \cite{BD2004}.
Let $\rho_1(x)$ and $H_1(x)$ be the solutions of { equations (\ref{eq-1},\ref{eq-3})}, in one dimension,  when point $A$ is at position $x=1$ and point $B$ 
is at position $x=0$. In this case, the  solution of  (\ref{Laplacian-1}) is simply
$$v_1 (x)=x$$
and $\rho_1$ and $H_1$ satisfy (\ref{eq-1},\ref{eq-3})
\begin{equation}
\Big( D(\rho_1(x))   \ \rho_1'(x) \Big)'=    \Big(    \sigma(\rho_1(x))  \,   H_1'(x)\Big)' 
\label{fq-1}
\end{equation}
\begin{equation}
D(\rho_1(x))  \   H_1''(x)=  -  {\sigma'(\rho_1(x) )  \over 2}   H_1'(x)^2 \, \, .
\label{fq-2}
\end{equation}

Then, using that $v(\vec{r})$ is solution of the Laplace equation (\ref{Laplacian-1}), it is easy to check that 
\begin{equation}
\label{sol-gene}
  H(\vec{r}) = H_1(v(\vec{r}))  \ \ \ ; \ \ \ 
  \rho(\vec{r}) = \rho_1(v(\vec{r})) 
\end{equation}
solve the equations (\ref{eq-1},\ref{eq-3}) { with the boundary conditions (\ref{HaHb},\ref{eq: bords rho})} in arbitrary dimension and for an arbitrary domain. {\it This is a central aspect of this work}. 
Replacing $H$ and $\rho$ by their expressions (\ref{sol-gene})  into (\ref{courant-2}) and subsequently into  
(\ref{mu-2})  leads to
\begin{eqnarray}
\label{mu-4}
\mu(\lambda)=  L^{d-2}
 \int d \vec{r} \  \left(\vec{\nabla} v(\vec{r}) \right)^2 \  \Big(  D(\rho_1)  \lambda   {\rho_1'} -   \\
  \lambda \, \sigma(\rho_1)  \,  H_1'   - \  { \sigma(\rho_1) \left(H_1' \right)^2 \over 2 } \Big) 
\nonumber
\end{eqnarray}
where $H_1$ and $\rho_1$  stand respectively for $H_1(       v(\vec{r}))$
and $\rho_1(       v(\vec{r})).$

Now the last step comes from the fact that for any function $E$ (any here means that $E$ is an arbitrary polynomial, and by extension any continuous function)
 the following identity holds
\begin{equation}
\label{ident}
 \int d \vec{r} \  
 \Phi(v(\vec{r}))  
\left(\vec{\nabla}        v(\vec{r}) \right)^2 
=  \int_0^1 \Phi(x) dx    \times
 \int d \vec{r} \ 
 \left(\vec{\nabla}        v(\vec{r}) \right)^2  
\end{equation}
whenever $v(\vec{r})$ satisfies
(\ref{Laplacian-1}) and (\ref{bound-cond}).
This identity is straightforward  when $\Phi(x)$ is constant. For a monomial $\Phi(x)=x^n$, it can be obtained using an integration by parts and (\ref{Laplacian-1},\ref{bound-cond}), namely,
\begin{eqnarray*}
 \int d \vec{r} \  
 v(\vec{r})^n  
\left(\vec{\nabla}        v(\vec{r}) \right)^2 
=  -\int d \vec{r} \  
 v(\vec{r})^n  
\vec{\nabla}(1-v(\vec{r})). \vec{\nabla}         v(\vec{r})
\\ =n \int d \vec{r} \  
( v(\vec{r})^{n-1} -   v(\vec{r})^n ) 
\left(\vec{\nabla}        v(\vec{r}) \right)^2 
\end{eqnarray*}
 so that
\begin{eqnarray*}
 \int d \vec{r} \  
 v(\vec{r})^n  
\left(\vec{\nabla}        v(\vec{r}) \right)^2 
= {n \over n+1} \int d \vec{r} \  
 v(\vec{r})^{n-1}  
\left(\vec{\nabla}        v(\vec{r}) \right)^2 
\\ = \cdots =   {1 \over n+1} \int d \vec{r} \  
\left(\vec{\nabla}        v(\vec{r}) \right)^2  \ , 
\end{eqnarray*}
thus showing that (\ref{ident})  should be true for any continuous  function $\Phi(x) $  as it can be approximated by a sum of monomials.

Applying the identity  (\ref{ident}) to (\ref{mu-4})  we deduce that 
\begin{eqnarray*}
\mu(\lambda) & =&  \left[ L^{d-2}
 \int d \vec{r} \  \left(\vec{\nabla} v(\vec{r}) \right)^2 \right]  \times   \int dx  \Big( \lambda  D(\rho_1(x) )  {\rho_1'(x)}\\  
& & \qquad \qquad -   
  \lambda \, \sigma(\rho_1(x))  \,  H_1'(x)   - \  { \sigma(\rho_1(x)) \left(H_1'(x) \right)^2 \over 2 } \Big) \\
& =&  \left[ L^{d-2}
 \int d \vec{r} \  \left(\vec{\nabla} v(\vec{r}) \right)^2 \right]  \times   \Big[ L  \mu_1(\lambda) \Big ]
\nonumber
\end{eqnarray*}
implying that $\mu(\lambda)$   for an arbitrary domain is the same as the one dimensional generating function $\mu_1(\lambda)$, up to a multiplicative  factor function independent of $\lambda, \rho_a, \rho_b$ and  of the functions $D$ and $\sigma$. 
Therefore the ratio between any pair of cumulants is the same as in one dimension.

\section{Conclusion}
In the present work we have shown that, at least,  whenever  the continuous macroscopic fluctuation theory can be applied, the cumulants  of the integrated current for the SSEP  
on an arbitrary large finite domain in dimension $d$ are the same as in one dimension. Our numerical   results of figure 3 for squares confirm this claim in  dimension $d=2$.
In dimension $d=3$, the discrepancy in figure 4 decreases with the size of the contacts and indicate that the continuous theory should become  applicable  for large enough contacts
  (the importance of the nature 
of connections was already pointed out in \cite{HOSS} in the quantum case).
 Our approach works as well for more general diffusive systems:  when $D(\rho)$ and $\sigma(\rho)$ are such that the optimal profiles are time independent the generating function $\mu(\lambda)$ should  be the same as in the one dimensional case.
Let us conclude by briefly mentioning { some open questions  related to the present work. 
(1)  Is it possible to generalize our results to the case where the optimal profiles become time dependent \cite{BDGJL5,BDGJL6,BD2005,BD2007} ? (2) What happens for an infinite  graph with a fixed density at infinity? (3) Is there an extension to  random graphs? (4) Is it possible to understand finite size corrections as in one dimension \cite{ADLV}? (5) Is there a generalization to systems in contact with more than 2  reservoirs at unequal densities? } 

\acknowledgments
We would like to thank Beno\^it Dou\c{c}ot for useful discussions. E.A and O.S acknowledge support by the Israel Science Foundation grant 924/09. The work of B.D and T.B was supported by the ANR-2010-BLAN-0108.

\end{document}